\documentclass[10pt,twocolumn]{ICCAS2021}
 
\usepackage[hyphens]{url}
\usepackage[hidelinks]{hyperref}
\hypersetup{breaklinks=true}
\usepackage{diagbox}
\usepackage{cite}
\usepackage{amsmath,amssymb,amsfonts}
\usepackage{algorithmic}
\usepackage{graphicx}
\usepackage{textcomp}
\usepackage{xcolor}

\begin{document}

\title{Development of an R-Mode Simulator Using MF DGNSS Signals}

\author{Suhui Jeong${}^{1}$ and Pyo-Woong Son${}^{2*}$ }

\affils{ ${}^{1}$School of Integrated Technology, Yonsei University, \\
Incheon, 21983, Korea (ssuhui@yonsei.ac.kr) \\
${}^{2}$Korea Research Institute of Ships and Ocean Engineering, \\
Daejeon, 34103, Korea (pwson@kriso.re.kr) \\
{\small${}^{*}$ Corresponding author}}


\abstract{
    With the development of positioning, navigation, and timing (PNT) information-based industries, PNT information is becoming increasingly important. Therefore, various navigation studies have been actively conducted to back up global positioning system (GPS) in scenarios in which it is disabled. Ranging using signals of opportunity (SoOP) has the advantage of infrastructure already being in place. Among them, the ranging mode (R-Mode) is a technology that uses available SoOPs such as a medium frequency (MF) differential global navigation satellite System (DGNSS) signal that has recently been recognized for its potential for navigation and is currently under research. In this study, we developed a signal simulator that considers the characteristics of MF DGNSS signals and skywaves used in R-Mode.
}


\maketitle


\section{Introduction}
Positioning, navigation, and timing (PNT) information is becoming an essential infrastructure in modern society that is required in various fields such as navigation, logistics, telecommunication, and finance. The most widely used PNT system today is the global navigation satellite system (GNSS) including the global positioning system (GPS) of the United States. Although GNSSs are widely used owing to their high accuracy and usability \cite{Enge1994:global, Park2020800, Kim20181087, Knoop17}, their weak signal strength makes them vulnerable to radio frequency interference \cite{Park2021919, Park2018387, Park20173888, Kim2019}. GNSSs are also vulnerable to ionospheric anomalies \cite{Chiou2008490, Seo20111963, Lee17:Monitoring, Sun2020889, Sun21:Markov, Ahmed20171792}. 
To overcome these limitations, research into the backup system is ongoing such as enhanced long-range navigation (eLoran) \cite{Son20191828, Son2018666, Williams13, Kim2020796, Qiu10, Park2020824, Hwang2018, Li20, Rhee21:Enhanced}, LTE-based positioning \cite{Shamaei21, Maaref20, Jeong2020958, Lee2020:Preliminary, Lee2020939, Jia21:Ground, Han2019, Lee20202347}, and other methods \cite{Rhee2019, Rhee2018224, Moon20171276, Kim2017:SFOL, Shin2017617, Kim2017348}. 
In particular, extensive research has been conducted on alternative navigation using ``signals of opportunity'' (SoOP) \cite{Mcellroy2006:navigation}. Navigation using SoOP means using radio frequency (RF) signals, whose original function was not navigation, for navigation, such as WiFi and LTE.

Medium frequency (MF) Differential GNSS (DGNSS) signals were originally aimed at increasing GNSS positioning accuracy by providing GNSS differential corrections to the user via an MF using minimum shift keying (MSK) modulation. As an SoOP, an MF DGNSS signal has the advantage of being widely distributed and its transmitter location known. 
Ranging mode (R-Mode) refers to an integrated terrestrial navigation system that uses various SoOPs such as a very high frequency data exchange system (VDES), eLoran, and MF DGNSS signal depending on the situation  \cite{Johnson2014:feasibility, Johnson:feasibility}. Especially, R-Mode technology using MF DGNSS signals has been actively studied \cite{Johnson:feasibility, Johnson2017:initial, Swaszek2012:ranging, Johnson2014:feasibility2}. One of the implementations of MF DGNSS R-Mode is to broadcast MF DGNSS signals by adding two continuous wave (CW) signals with a 500 Hz bandwidth onto the MSK carrier for ranging \cite{Johnson:feasibility}.

Since R-Mode techniques using MF DGNSS are currently being studied, simulators are required to predict and improve the performance. The largest source of error for MF R-Mode is the skywave \cite{Johnson:feasibility}. In this study, considering the transmission and reception environment, we developed a software simulator that generates a received signal by combining an MF DGNSS signal, two CW signals, and a skywave reflected by the ionosphere.

\section{MF DGNSS Signal Model}

DGNSS is a system in which reference stations on the ground broadcast differential correction messages to nearby users to increase the GNSS accuracy. The differential corrections are generated based on the difference between the measured range and actual range between the satellite and reference station.

The MF DGNSS system uses an MF band of 285--325 kHz and transmits differential correction information using MSK modulation. MSK modulation has the advantage of not requiring to consider the amplitude at the receiving end because it has a constant amplitude \cite{Swaszek2012:ranging}. When transmitting the \textit{k}-th data bit, MSK modulated signals can be represented by the following expression \cite{Swaszek2012:ranging}.
\begin{equation} 
  \label{eqn:msk}
  \begin{split}
    {s_{\mathrm msk}} \left ( t \right ) 
      = \left\{
        \begin{matrix} A\sin \left ( 2\pi f_{0}t+\phi _{k} \right ); 
        k \textrm{-th data bit} = 0\\
        A\sin\left ( 2\pi f_{1}t+\phi _{k} \right ); 
        k \textrm{-th data bit} = 1 
        \end{matrix}\right.
  \end{split}
\end{equation}
where $A$ is the amplitude of signals,  $f_{0}$ and $f_{1}$ are two frequencies that satisfy $\left | f_{0} - f_{1} \right | = 1/(2T)$, $1/T$ is the data rate with a value of 100 or 200 bits per second, $t$ is time, and $\phi _{k}$ is a phase that enables MSK signals to have a continuous phase. 

We can rewrite Eq. (\ref{eqn:msk}) 
as \cite{Pasupathy1979:minimum}:
\begin{equation} 
  \label{eqn:msk_re}
  \begin{split}
    s_{\mathrm msk} =& \pm b_{\mathrm msk}\cos \left ( \frac{\pi t}{2T} \right )\cos 2\pi f_{c}t \\
    & \pm b_{\mathrm msk}\sin \left ( \frac{\pi t}{2T}\right )\sin 2\pi f_{c}t\\ 
    =& \pm b_{\mathrm msk}\cos \left [ 2\pi f_{c}t+d_{k}\frac{\pi t}{2T} \right ]\\ 
    =&  b_{\mathrm msk}\cos \left [ 2\pi f_{c}t + d_{k}\frac{\pi t}{2T}+\Phi _{k} \right ] 
  \end{split}
\end{equation}
with
\begin{equation*} 
  \begin{split}
    &d_{k}=-I_{k}Q_{k } \\
    &\Phi _{k}=\frac{\pi }{2}\left ( 1-I_{k} \right )
  \end{split}
\end{equation*}
where $b_{\mathrm msk}$ is the amplitude factor, $f_{c}$ is the carrier frequency, and $I_{k}$ and $Q_{k}$ are the inphase and quadrature channel's \textit{k}-th data bit, respectively.

One of the implementations of the MF DGNSS R-Mode adds two CW signals to the DGNSS signal at $f_{c} \pm 250$ Hz for ranging. CW signals help facilitate phase measurement, and a combination of the two CW signals aids in resolving  ambiguity \cite{Johnson:feasibility}. The CW signals $s_{\mathrm cw1}$ and $s_{\mathrm cw2}$ are expressed as follows ($f_c$ is in Hz.):
\begin{equation} 
    \label{eqn:CW}
  \begin{split}
    &s_{\mathrm cw1} \left( t \right) = b_{\mathrm cw1} \sin \left( 2\pi \left( f_{c} - 250 \right) t +\Phi_{\mathrm cw1} \right)\\
    &s_{\mathrm cw2} \left ( t \right) = b_{\mathrm cw2} \sin \left( 2\pi \left( f_{c} + 250 \right) t +\Phi_{\mathrm cw2} \right).
 \end{split}
\end{equation}
The total transmitted signal is represented by the sum of $s_{\mathrm msk}$, $s_{\mathrm cw1}$, and $s_{\mathrm cw2}$.
\begin{equation} 
\label{eqn:TotalSignal}
  \begin{split}
     s\left ( t \right ) =& s_{\mathrm msk}\left ( t \right ) + s_{\mathrm cw1}\left ( t \right ) + s_{\mathrm cw2}\left ( t \right )\\
    =& b_{\mathrm msk}\cos \left ( 2\pi f_{c}t + d_{k}\frac{\pi}{2T}t+\Phi _{k} \right )\\
    &+ b_{\mathrm cw1}\sin \left ( 2\pi\left ( f_{c}- 250 \right ) t +\Phi _{\mathrm cw1} \right )\\
    &+ b_{\mathrm cw2}\sin \left ( 2\pi\left ( f_{c}+ 250 \right ) t +\Phi _{\mathrm cw2} \right ).
 \end{split}
\end{equation}
Note that the received signals of the user are different from Eq. (\ref{eqn:TotalSignal}) because they are affected by skywaves.

\section{Skywave Model}

\begin{figure}
    \centering
    \includegraphics[width=0.9\linewidth]{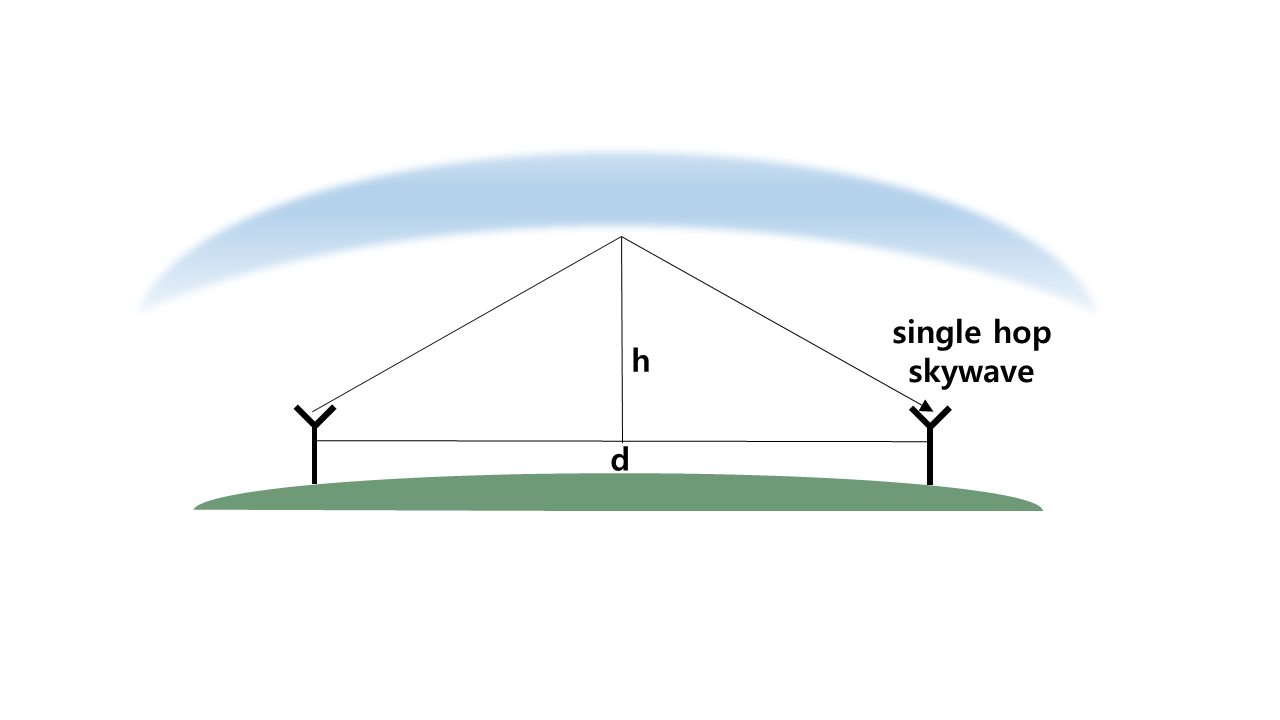}
    \caption{Single hop skywave}
    \label{fig:skywave}
\end{figure}

A skywave is a signal in which an original signal is reflected by the ionosphere. MF DGNSS ground wave and skywave signals are received together. A skywave that is reflected once by the ionosphere with distance $d$ in Fig.~\ref{fig:skywave} is expressed as follows \cite{Johnson:feasibility}:
\begin{equation} 
 \label{eqn:skywave}
    r\left ( t \right )=s\left ( t \right )+\alpha s\left ( t-t_{d} \right ); \:\:\: 0<t<T
\end{equation}
where $\alpha$ is the attenuation factor, and $t_{d}$ is the travel time difference of the groundwave and skywave, which can be represented by subtracting the skywave propagation distance with groundwave propagation distance divided by the speed of light. If the ionospheric height, where the reflection occurs, is $h$, as in Fig.~\ref{fig:skywave}, $t_{d}$ can be represented as follows:
\begin{equation} 
 \label{eqn:td}
    t_{d}=\frac{\sqrt{4 h^2 + d^2}-d}{c}
 \end{equation}
where $c$ is the speed of light.
In other words, the skywave is a time-delayed and attenuated signal from the original signal.

The characteristics of MF skywave are that, first, the travel time of the signal is longer than that of the ground wave. Second, because of changes in the height of the ionosphere, the effect of the skywave is greater at night than during the day \cite{Hunsucker2007:high}. 
Finally, the greater the distance, the greater will be the effect of the skywave \cite{Johnson:feasibility}.

Johnson \textit{et al.}  \cite{Johnson:feasibility} analyzed the effect of skywaves on the localization performance of R-Mode. When ranging with the MSK signal, it is important to know the phase and bit transition time of the signal. However, skywave signals cause changes in the signal phase and the bit transition time. 
Ignoring noise, assuming that the transmitted signal is a simple sine wave, the received signal, including the ground wave and skywave, is expressed as follows \cite{Johnson:feasibility}:
\begin{equation} 
 \label{eqn:skywave_re}
 \begin{split}
    r\left ( t \right )
    &= s\left ( t \right )+\alpha s\left ( t-t_{d} \right )\\ 
    &= B\sin \left ( \omega _{k}t+\phi \right )+\alpha B\sin \left ( \omega _{k}\left ( t-t_{d}\right)+\phi \right )\\
    &= \eta B\sin \left ( \omega _{k}t+\phi +\beta \right) 
\end{split}
\end{equation}
where $B$ is the amplitude of $s(t)$, $\eta$ is the amplitude scaling, and $\beta$ is the phase shift, which expressed as follows:
\begin{equation} 
\label{eqn:coefficient}
\begin{split}
    &\eta =\sqrt{1+\alpha ^{2}-2\alpha\cos \left ( \omega _{k}t_{d} \right )}\\
    &\beta =\tan^{-1}\left ( \frac{\alpha \sin \left ( \omega _{k}t_{d} \right )}{1-\alpha \cos \left ( \omega _{k}t_{d} \right )} \right )
\end{split}
\end{equation}

As the amplitude changes, the bit transition time becomes unclear, and the phase also changes. Eq. (\ref{eqn:coefficient}) shows that a small $\alpha$ can reduce the size of $\eta$, but the phase shift can vary significantly with $t_{d}$. The effects of skywaves during the day are not significant and those of night time are significant. Therefore, when creating an R-Mode simulator, the effect of the skywave should be fully considered.

\section{Software configuration}
\begin{figure}
    \centering
    \includegraphics[width=0.9\linewidth]{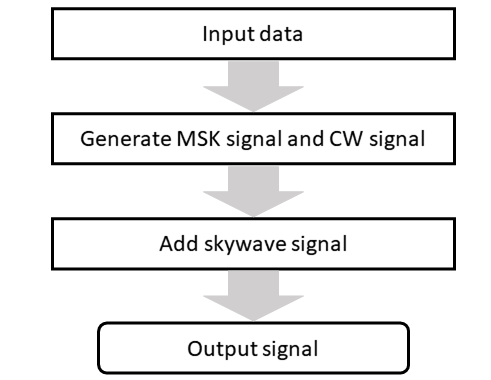}
    \caption{Process of the software}
    \label{fig:ProcessOftheSoftware}
\end{figure}

The aim of the developed software is to create a received signal for R-Mode using MF DGNSS. The functionality of this software is as follows. 
First, the transmission signal information considers the signal transmit power, carrier frequency, and transmit data rate. Based on these data, MSK-modulated signals are generated, and two CW signals are added. 
Second, the skywave latency is calculated based on the height of the reflected ionospheric layer and the distance between the transmitter and receiver. 
The attenuation factor of the skywave is calculated using the software named ``Calculation of LF-MF field strength and phase with parabolic model'' \cite{ITU_LFMF} provided by the international telecommunication union (ITU). 
A skywave is created and added considering attenuation and latency. 
Finally, additive white Gaussian noise (AWGN) is added to generate a received signal. 
Fig.~\ref{fig:ProcessOftheSoftware} shows a block diagram of this process.

\section{Implementation Result} 

\begin{figure}
    \centering
    \includegraphics[width=0.9\linewidth]{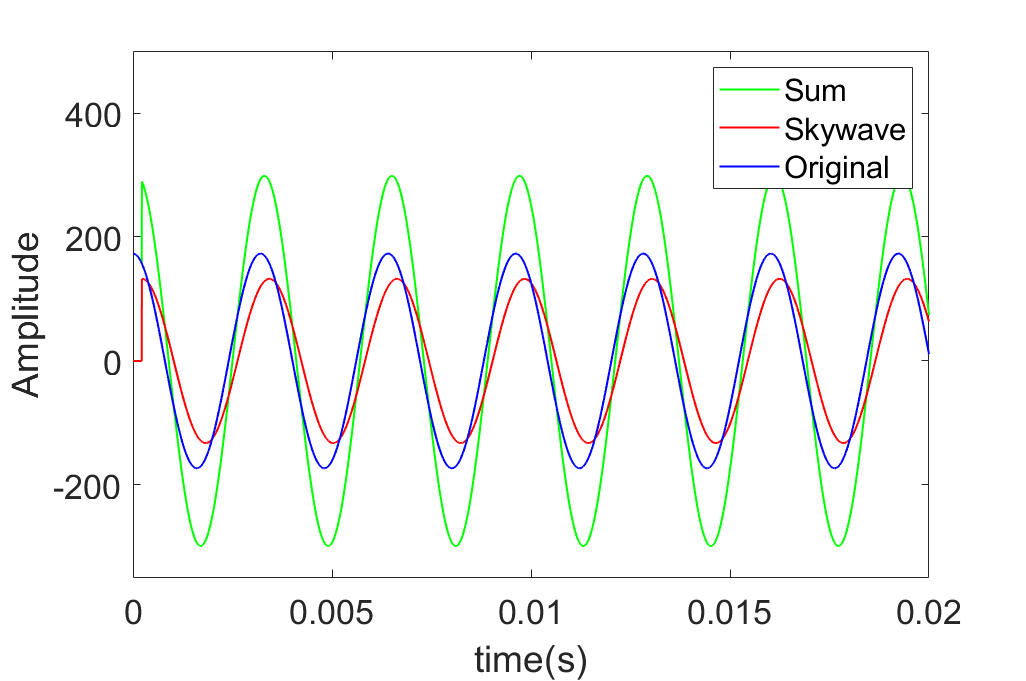}
    \caption{Simulation result}
    \label{fig:SimulationResult}
\end{figure}

Fig.~\ref{fig:SimulationResult} is the result of a plot of signals to be received at a distance of 210 km using MF DGNSS transmission signals from Geomundo Island in South Korea. It produced transmission signals under the conditions that transmit power at 150 kW, carrier frequency, $f_c$, at 287 kHz, and data rate, $1/T$, at 100 bps, with the addition of CW signals at $f_c \pm 250$ Hz. The ionosphere height, $h$, used for skywave time delay calculations was 90 km, and the attenuation factor, $\alpha$, was obtained using the ITU software. The blue line is the sum of the MSK and CW signals, the red line is the skywave signal, and the green line is the sum of the two.

\section{Conclusion}

In this study, we created a simulator for R-Mode studies using MF DGNSS signals. Considering the transmission and reception environment, MF DGNSS signals, CW signals, and skywaves were modeled and implemented together by substituting the actual environmental conditions. In the future, based on the software created in this study, we will implement receiver parts that demodulate received signals, and further considerations will be made in the transmission and reception environment. Later, the software will be used for various studies to improve the R-Mode performance.

\section*{ACKNOWLEDGEMENT}

This research was supported by a grant from National R\&D Project ``Development of integrated R-Mode navigation system'' funded by the Ministry of Oceans and Fisheries, Korea.

\bibliographystyle{IEEEtran}
\bibliography{mybibfile, IUS_publications}

\end{document}